# FROM SERVICE QUALITY TO E-SERVICE QUALITY: MEASUREMENT, DIMENSIONS AND MODEL

Ighomereho Ogheneochuko Salome, Redeemer's University
Ojo Afolabi Ayotunde, Redeemer's University
Omoyele Olufemi Samuel, Redeemer's University
Olabode Oluwayinka Samuel, Redeemer's University

**ABSTRACT**

*With the global increase in online services, there is a paradigm shift from service quality to e-service quality. In order to sustain this strategic change, there is need to measure and evaluate the quality of e-services. Consequently, the paper seeks to determine the relevant e-service quality dimensions for e-channels. The aim is to generate a concise set of dimensions that managers can use to measure e-service quality. The paper proposed an e-service quality model comprising seven e-service quality dimensions (website appearance, ease of use, reliability, security, personalisation, fulfilment and responsiveness) and overall e-service quality. The study employed a cross-sectional research design and quantitative research approach. The data were collected via a questionnaire from 400 e-channel users in Lagos State, Nigeria. However, 318 copies of the questionnaire were found useful. The data were analysed using mean, frequency, percentages, correlation and multiple regression analysis. The results revealed that the relevant e-service quality dimensions influencing overall e-service quality are reliability, security, fulfilment, ease of use and responsiveness. These e-service quality dimensions are expected to provide information for managers to evaluate and improve their e-channel service delivery.*

## INTRODUCTION

The service sector is an important segment of all economies and an essential part of our daily lives (Ejigu, 2016). Hence, the quality of services is a major concern in the management of services. Service quality describes a service that fulfills customers' expectations and satisfies their needs (Sadaf & Rahela, 2019). It can also be described as the subjective comparison that customers make between the quality expectations of a service and what they receive. It is, therefore, a form of attitude, which emanates from the comparison of expectations with actual performance. It has been noted that customers' perception of service quality depends on customers' pre-service expectations (Zeithaml, Parasuraman & Malhotra, 2000). Customers will judge the quality of a service as low if the performance does not meet their expectations and high when the performance exceeds expectations (Jayaraman, Shankar & Hor, 2010). Managers and academics are keen on accurately measuring service quality to understand its consequences and to establish techniques for improving quality in order to achieve customer satisfaction and to minimise customer switching behaviour (Mohammad, Asaad & Ihab, 2018).

Empirical evidence indicates that the success and sustainability of firms depend largely on the quality of services. Several authors (Biljana and Jusuf, 2011; Gong and Yoo, 2013; Taherikia and Shamsi, 2014; Zehir and Narcıkara, 2016; Al-hawary and Al-Smeran, 2017; Singh, 2019; Nihayah, Marwan, Nawras, Ahmad and Ahmad, 2021) have noted the relevance of service quality to a business and have highlighted its contribution to competitive advantage, profits, increased markets share, return on investment, customer satisfaction and future purchasing intention. Parasuraman, Zeithaml and Malhorta (2005) opined that the key strategy for the success and survival of any business is to deliver quality services to the target market. When firms provide high service quality, it increases efficiency in service delivery, leading to an increase in business's profitability. In addition, the provision of quality services might result in repeated purchases and extended positive word of mouth (Taherikia & Shamsi, 2014). These roles of service quality in business performance have led to the search for dimensions to measure service quality.

The dimensions for measuring service quality have been a significant discourse in marketing research (Riadh, 2008; Yarimoglu, 2014; Nihayah et al., 2021). Majority of the service quality research has been devoted to designing and developing service quality measures. Several researchers (Gronroos 1984; Parasuraman et al., 1985; 1988; Zeithaml et al., 2002, Narteh, 2013) have attempted to articulate and design models for determining the dimensions of service quality. For e-service quality, there is yet to be a concise set of dimensions. With the rapid growth of online services, there is need to develop measures that will assist researchers and practitioners in measuring the dimensions of e-service quality that are relevant to e-channels. In Nigeria, the adoption of e-services is growing for online shopping, e-banking, e-retailing, e-commerce, e-ticketing, e-booking, e-learning, e-insurance, among others. This is expected to increase globally due to the benefits of e-channels. However, some researchers (Bauer et al., 2006; Olasanmi, 2019; Mustapha et al., 2021; Anene & Okeji, 2021) have noted that there are issues and challenges with e-service quality. Hence, the objectives of the paper are to propose a model for the measurement of e-services and to determine the e-service quality dimensions that influence overall e-service quality.

## LITERATURE REVIEW

### Service Quality Dimensions

Service quality dimensions are a set of features that describe customers' experience with a service. Some service quality features have been propounded to explain the dimensions that influence customers' perception of service quality. The primary goal of the dimensions is to offer managers and researchers insights into the dimensions of service quality that can improve service offerings. Extant literature examined service quality dimensions from two perspectives, namely, traditional service quality and e-service quality.

### Traditional Service Quality

As defined by Zeithaml et al. (2000) "traditional service quality relates to the quality of all non-technology-based customer interactions and experiences with firms". The dimension for measuring service quality in this setting was a challenge to service providers because services have many characteristics that distinguish them from physical goods (Sadaf & Rahela, 2019). Services can be described as processes that are intangible and heterogeneous and so they cannot be kept as inventory. Moreover, there is no transfer of ownership. Furthermore, the production, distribution, and consumption take place simultaneously, and most importantly, customers are involved in the production process. This last attribute of services has significant implications when service quality is discussed. This is due to the involvement of customers in the process of production. As customers receive the service, the quality is directly perceived by the customers.

Some researchers have made attempt to identify the features of a service that are essential to quality evaluations. For example, Grönroos (1984) measured service quality dimensions based on functional quality and technical quality. Parasuraman et al. (1985) identified ten dimensions that formed the basis of customer evaluation of service quality. They are reliability, responsiveness, competence, access, courtesy, communication, credibility, security, understanding/knowing the customer and tangibles. However, Parasuraman et al. (1988) conducted empirical studies in several industry sectors to develop and refine service quality dimensions and quantify customers' global (as opposed to transaction-specific) assessment of a company's service quality. Based on scale enhancement, the initial ten dimensions were condensed to five dimensions of tangibles, reliability, responsiveness, assurance and empathy, usually referred to as SERVQUAL.

Cronin and Taylor (1992) developed the SERVPERF model to measure service quality based on customer's overall feeling towards the service. Rust and Oliver (1994) proposed three dimensions: service product, service delivery, and service environment. Dabholkar, Thorpe and Rentz (1996) developed an empirically validated multilevel model called Retail Service Quality Scale (RSQS) with five dimensions: physical aspects, reliability, personal interaction, problem solving, and policy. Philip and Hazlett (1997) developed a hierarchical structure model called PCP for measuring service quality. The model was based on pivotal, core and peripheral attributes. Pivotal attributes which are the most criticalattributes that affect service quality, were seen as end product or output, whereas; core and peripheral attributes were seen as inputs and processes. Brady and Cronin (2001) identified service quality dimensions as interaction quality, physical service environment quality, and outcome quality.

Of all the service quality dimensions, the most extensively used dimension of traditional service quality is the SERVQUAL model propounded by Parasuraman et al. (1988). The five dimensions of SERVQUAL were found to be valid in the traditional service environment; however, they have been found unsuitable for determining the quality of services delivered over the web and other technology-based services. Despite the application of SERVQUAL model for determining e-service quality by some researchers (Phan & Nham, 2015; Mwatsika, 2016), some other researchers (Hongxiu, Yong & Reima, 2009; Barrutia & Gilsanz, 2009; Tan, Abdul, Zahir, Lee, Arif, Parameswaran & Rasheedul, 2018) have raised some concerns about it. This is because the service delivery processes of e-services differ significantly from that of traditional service. E-service is different from traditional service based on two major attributes of e-service, which are the absence of staff and self- service by customers. Therefore, the attributes for defining a high quality service are expected to differ in the two contexts. This was noted by Yang and Fang (2004) when they stated that some of the service quality dimensions relevant to traditional services might not berelevant for online services. This led to the search for the dimensions of e-service quality.

**E-Service Quality**

From the year 2000, e-service practices have been on the increase. The concept of "e-service" emerged from the growth of the internet and its application in business. This development brought about e-service quality, and several authors have offered a variety of definitions. Zeithaml et al. (2000) defined e-service quality as the extent to which a website facilitates efficient and effective shopping, purchasing and delivery of services. In the view of Parasuraman et al. (2005) e-service quality involves all phases of a customer's interactions with a website. To further understand the nature of e-services, Zeithaml et al. (2002)reviewed the gap model of service quality and proposed the gap model of e-service quality, which is a refinement of the earlier and well-accepted gap model of service quality. This change became necessary because e-service is delivered to customers through a technology, and there was a need to incorporate the human-technology interaction.

Several researchers have focused on conceptualising and measuring e-service quality and examining its effects in the electronic marketplace. They tried to identify the dimensions of e-service quality by measuring it with the help of specific attributes of the given service, as indicated in Table 1.

**TABLE 1**
**DIMENSIONS OF E-SERVICE QUALITY**

| S/N | Author(s) | Year | Study | Dimensions of E-Service Quality Used |
|---|---|---|---|---|
| 1 | Zeithaml et al. | 2000 | Online Shopping | Reliability, Responsiveness, Access, Flexibility, Ease of navigation, Efficiency, Assurance/trust, Security/privacy, Price knowledge, Site aesthetics and Customisation/personalisation. |
| 2 | Santos | 2003 | E-Commerce | Incubative dimensions: Web appearance, Ease of use, Linkage layout, Content Active dimensions: Reliability, Efficiency, Security, Support, Communication, Incentive |
| 3 | Bauer et al. | 2006 | Online Shopping | Functionality/design, Reliability, Process, Responsiveness, Enjoyment. |
| 4 | Hongxiu et al. | 2009 | Online travel service | Ease of use, Website design, Reliability, Service availability, Privacy, Responsiveness, Empathy |
| 5 | Alsudairi | 2012 | E-Banking | Accessibility, Usability, Functional usefulness, Safety, Convenience, Responsiveness, Realisation. |
| 6 | Narteh | 2015 | ATM | Convenience, Reliability, Ease of Use, Privacy and Security, Responsiveness, Fulfilment |
| 7 | Tan et al. | 2018 | Internet Retail Service | Information, Navigation, Security, Responsiveness, Reliability. |
| 8 | Nihayah et al. | 2021 | E-Banking | Accessibility, Responsiveness, Functional usefulness, Usability, Safety, Convenience, Realisation |

Table 1 shows that several dimensions of e-service quality have been developed over the past several years for specific e-service. The literature indicates that the studies have been conducted mainly in the area of online shopping, e-commerce, online retailing, online travel service and e-banking. This implies that there is no agreement among researchers concerning the important dimensions of e-service quality. However, some dimensions appear to be common. So, there is need to develop a concise set of dimensions that can be applied to measure e-service quality irrespective of the type of e-channel.

## METHODS

To achieve the objectives of the study, a cross-sectional research design and quantitative survey research approach was adopted. The literature on e-service quality was subjected to content analysis, resulting in the identification of fifteen e-service quality dimensions (website appearance, ease of use, access, efficiency, reliability, assurance, trust, fulfilment, responsiveness, personalisation, communication, navigation, service availability, content and security). A list of these dimensions was presented to thirty (30) e-channel users in Lagos State, Nigeria, to determine the dimensions for this study on a five-point Likert scale ranging from very important to very unimportant. The aim was to ensure that only e-service quality dimensions found in the literature that are important to e-channel users were tested. Seven of the e-service quality dimensions (website appearance, ease of use, reliability, security, personalisation, fulfilment and responsiveness) were highly rated and were used for the study. The target respondents are e-channel users who have a minimum of 6 months usage. Hence, the sample was selected using purposive sampling technique. A questionnaire incorporating the seven dimensions was used to collect data to identify the relevant e-service quality dimensions. The questionnaire consists of 50 items categorised into three parts. The first part comprises demographic information and usage behaviour; the second part consists of e-service quality dimensions, while the third part addresses overall e-service quality. 400 copies of the survey questionnaire were distributed to e-channel users in Lagos State, Nigeria. To ensure proper distribution of the questionnaire, 100 copies each were administered to students, private employees, civil servants and self-employed. 336 copies were retrieved, but 318 were completely filled. Therefore, 318, giving a response rate of 79.5%, were used for the analyses.

# MODEL SPECIFICATION

**Proposed Model**

Several dimensions of e-service quality have been propounded to improve service offerings. Drawing on the literature, fifteen (15) e-service quality dimensions were identified. Table 2 shows the results of the importance rating of the e-service quality dimensions by 30 e-channel users.

**TABLE 2**
**DESCRIPTIVE STATISTICS OF IMPORTANCE RATING OF E-SERVICE QUALITY DIMENSIONS (N=30)**

| E-Service Quality Dimension | N | Minimum | Maximum | Mean | Std. Deviation |
|---|---|---|---|---|---|
| Security | 30 | 5.00 | 5.00 | 5.00 | 0.00 |
| Ease of Use | 30 | 4.00 | 5.00 | 4.86 | 0.34 |
| Fulfilment | 30 | 4.00 | 5.00 | 4.83 | 0.37 |
| Reliability | 30 | 4.00 | 5.00 | 4.76 | 0.43 |
| Responsiveness | 30 | 3.00 | 5.00 | 4.40 | 0.72 |
| Website Appearance | 30 | 3.00 | 5.00 | 4.33 | 0.75 |
| Personalisation | 30 | 3.00 | 5.00 | 4.03 | 0.74 |
| Trust | 30 | 3.00 | 5.00 | 3.93 | 0.63 |
| Support | 30 | 3.00 | 5.00 | 3.80 | 0.84 |
| Assurance | 30 | 3.00 | 5.00 | 3.73 | 0.73 |
| Communication | 30 | 3.00 | 5.00 | 3.66 | 0.80 |
| Access | 30 | 3.00 | 5.00 | 3.60 | 0.77 |
| Navigation | 30 | 3.00 | 5.00 | 3.60 | 0.67 |
| Efficiency | 30 | 3.00 | 5.00 | 3.60 | 0.56 |
| Content | 30 | 2.00 | 4.00 | 3.03 | 0.55 |

Source: Survey (2021)

As indicated in Table 2, apart from content with a minimum response of 2, all the other dimensions have a minimum and maximum response of average (3) and above. This indicates that the respondents considered most of the dimensions to be important. For the purpose of this research, only the variables with a mean above 4.00 point were selected. Consequently, the paper proposed seven e-service quality dimensions in the context of e-channels. They are security, ease of use, fulfilment, reliability, responsiveness, website appearance and personalisation. The selected e-service quality dimensions are depicted in Figure 1.

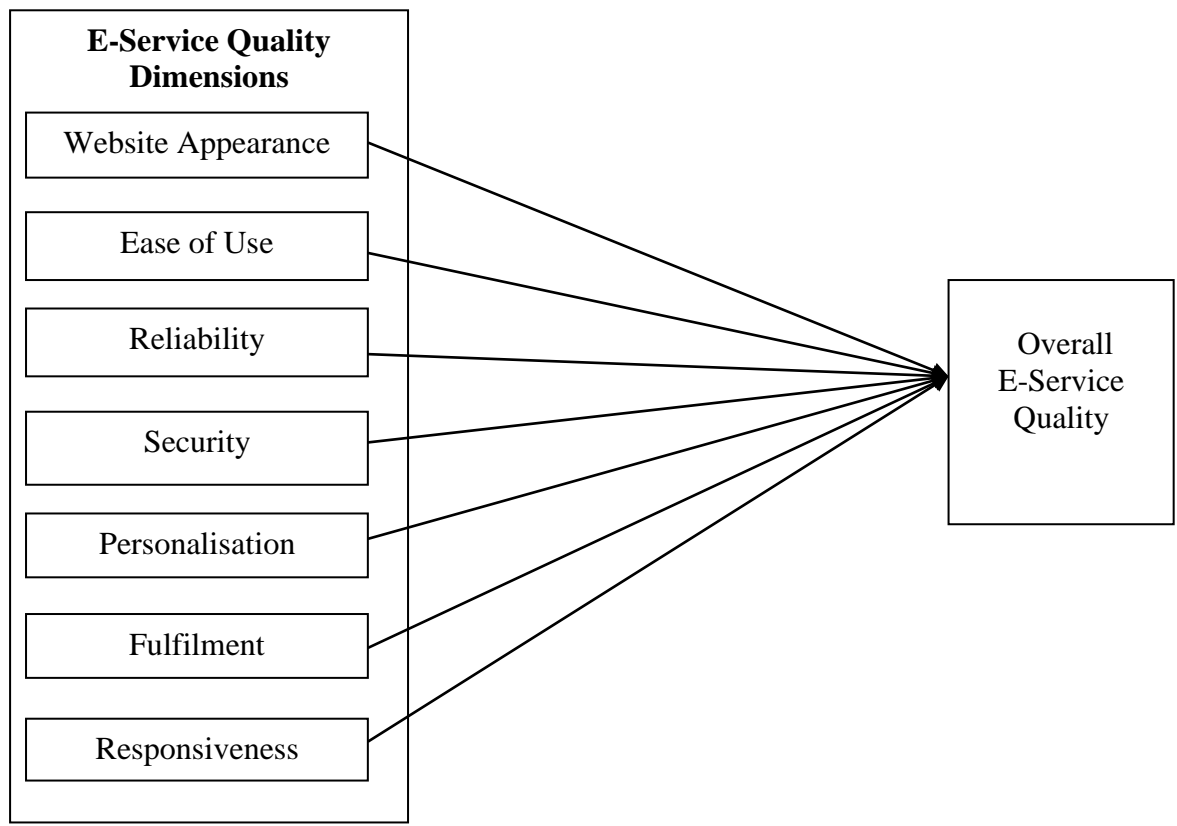

**FIGURE 1**

**MODEL OF E-SERVICE QUALITY**

Mathematically, the model is depicted as:

$$Y = a + b_1 X_1 + b_2 X_2 + b_3 X_3 + b_4 X_4 + b_5 X_5 + b_6 X_6 + b_7 X_7 + e \qquad (1)$$

Where,
Y = Overall E-Service Quality; a = Constant; $b_1$ to $b_7$ = Regression Coefficients for $X_1$-$X_7$; $X_1$ = Website Appearance; $X_2$ = Ease of Use; $X_3$ = Reliability; $X_4$ = Security; $X_5$ = Personalisation; $X_6$ = Fulfilment; $X_7$ = Responsiveness; e = error term

**Overall E-Service Quality**

Khushdil (2018) defined overall, e-service quality as customers' overall assessment of the utility of a service on the basis of the perceptions of what is received and what is given.

In the view of Zeithaml et al. (2000) "it is customers' judgment about an entity's overall excellence or superiority". It results from customers' comparison of their perceptions about the service delivery process and the actual outcome of a service (Lovelock & Wirtz, 2011). According to Narteh (2013) overall e-service quality portrays a general, overall appraisal of a service. It is the evaluation of the service performance customers received according to whether it meets certain standards.

**E-Service Quality Dimensions**

**Website appearance**

Website appearance is how the site looks, and it includes the site aesthetics, information structure, colour, animation, pictures, text, format, sound and visual design (Poon & Lee, 2012). As noted by Taherikia and Shamsi (2014) all items on a website should be explained in simple language so that it is clear to most users. According to Lee and Lin (2005) the user interface should be visually appealing and tidy to attract customers. Based on previous studies, website appearance plays a significant role in how customers judge service quality (Cox & Dale, 2001; Santos, 2003; Lee & Lin, 2005; Hongxiu et al., 2009; Khushdil, 2018; Paulo, Tiago & Almira, 2019). The website is the starting point for online customers to gain access to the business organisation. It is possible that it can influence customers' perception of a business and attract customers to engage in online services. As such, the websites should be well designed to be visually appealing. However, Raman, Stephenaus, Alam and Kuppusamy (2008) opined that a website with many flash animations, pop-up advertisements and graphic banners would dissatisfy the user. They noted that online customers are only interested in engaging in whatever they want to do rather than seeing animations.

**Ease of Use**

This is the degree to which the e-channel can be understood and operated in a simple and easy way. It also refers to the ability of a customer to find information or enact a transaction with the least amount of effort. It has been found to be one of the main determinants of e-service quality (Shirshendu & Sanjit, 2011; Narteh, 2013; Narteh, 2015). However, Yazeed et al. (2014) noted that some users find the instructions on how to perform some operations quite challenging to understand. So, if users feel that a system is easy to use, then the chances of using the system will be greater. A user-friendly e-channel may be important in increasing customers' utility which increases the probability of obtaining loyal customers. On the other hand, when the e-channel is not easy to use, it may prevent users from finalising the desired transaction and consequently, the users may not revisit the e-channel (Al-Hawari, Hartley & Ward, 2006).

**Reliability**

Reliability relates to accuracy, speed and constant availability of a service (Muhammad et al., 2014). In the opinion of Narteh (2015) reliability is the ability of the online platform to perform the promised service dependably, consistently and accurately. It also means that the business honours its promises. It involves accuracy in billing, keeping records correctly and performing the service whenever there is a request. Reliability means that the site functions all the time and is available 24/7 as promised. In e-service quality research, reliability has been found to be the most significant determinant of e-service quality and customer satisfaction (Narteh, 2013; Mwatsika, 2016; Al-Hawary & Al-Smeran, 2017; Tan et al., 2018). The importance of reliability is based on the premise that customers' perception of e-service quality is likely to increase when the service is performed as promised or expected. If customers cannot use the e-channel when they need the service, they may deviate from using it.

**Security**

Narteh (2013) defined security as the protection of customers from fraud and financial loss as well as the protection of customers' personal information. Although security concerns differ among countries, e-channels can be accessed globally, and so, Lee & Lin (2005) advised that online channels should provide secure transactions to make customers feel comfortable when using it. Zhengwei and Jinkun (2012) asserted that security holds a vital position in e-service because customers perceive significant risks in the virtual market space due to the high prevalence of internet fraud. Customers' perception of risks tends to be high for online services, especially financial services, because customers believe that the internet payment channels are not secure and can be intercepted, reducing the customers' trust level. This tends to discourage them from engaging in online information search and making online banking transactions. According to Agbonifoh et al. (2007) two kinds of security are desired by customers who use the internet, namely, informational and transactional securities. Informational security is associated with safety from loss arising from unauthorised persons' illegal use of customers' information. In contrast, transactional security refers to safety over business deals carried out over the internet. Security has been found to be an essential dimension of e-service quality (Madu & Madu, 2002; Wolfinbarger & Gilly, 2003; Akinmayowa & Ogbeide, 2014; Paulo et al., 2019) however; Narteh (2013) did not found security as a significant factor.

**Personalisation**

Lee and Lin (2005) defined personalisation as customer perception of the individualised attention and differentiated service tailored to meet individual needs and preferences. It is the ability of the online channel to address users on a one-on-one basis. It involves an understanding of customer needs, preferences and expectations and addressing them in the web. In this way, the content of the website will give the feeling that it is specifically designed for the customers. It should also acknowledge repeat customers by their names whenever they log in to the website. This provides customers with the feeling of familiarity and closeness, thus positively influencing customer relationship and customer loyalty (Kaynama & Black, 2000; Poon & Lee, 2012). As noted by Ojasalo (2010) personalisation may be done based on past purchases and other information provided by customers. E-service enables a business to collect and store information about its customers and identify them individually. When the customer database is linked to the website, the business can greet them with targeted offers whenever they visit the site. The more they buy or receive the service online, the more the company can effectively refine their profile and market to them.

**Fulfilment**

Fulfilment is the extent to which an e-channel performs outcomes that meet customer's expectations in terms of the extent to which the site's promises about order delivery and item availability are fulfilled (Ojasalo, 2010). It represents the outcome of performance of service delivery, and the focus is on customers' requirements in terms of the purpose for using the e-channel and what they receive. As noted by Ahmed, Romeika, Kauliene, Streimikis and Dapkus (2020) fulfilment examines the implementation of website promises. Fulfilment also involves the accurate display and description of a product so that what customers receive should be what they ordered for and the delivery of the right product within the time frame promised. Some authors (Sakhaei, Afshari and Esmaili, 2014; Narteh, 2015; Paulo et al., 2019) have found fulfilment to be a determinant of overall e-service quality.

**Responsiveness**

Responsiveness means speedy handling of problems and returns through the site (Tan et al., 2018). It has to do with how customer care or support service responds to help customers when they face problems with a service (Muhammad et al., 2014). It is the business ability to handle customer complaints due to transactional failures. It includes the extent to which the business has put measures to recover services when the e-channel could not deliver as expected. E-channel users expect quick feedback on requests. So, it also involves the attention and promptness in dealing with customer requests, questions, complaints, and issues and compensating customers when they encounter financial losses. Therefore, the ability to handle customer questions, concerns, and frustrations is essential to the perception of e-service quality. Madu and Madu (2002); Narteh, (2015); Al-Hawary and Al-Smeran (2017); Tan et al., (2018) found responsiveness to be a critical factor in e-service quality.

## RESULTS

**Demographic Characteristics of the Respondents**

The demographic characteristics of the respondents indicate that (171) 53.8 per cent of the respondents were males while (147) 46.2 per cent were females. 54 of the respondents 17.0 per cent were below 21 years, (109) 34.3 per cent were between 21-30 years, (92) 28.9 per cent were between 31-40 years, (53) 16.7 per cent were between 41-50 years and (10) 3.1 per cent were 51 years and above. For the level of education of the respondents, (15) 4.7 per cent had primary education, (100) 31.4 per cent had secondary education, (161) 50.6 per cent had tertiary education, while (42) 13.2 per cent had professional education. In terms of employment status, (88) 27.7 per cent were students, (84) 26.4 per cent were private employees, (86) 27.0 per cent were civil servants, while (60) 18.9 per cent were self-employed. With respect to monthly income/allowance, (156) 49.1 percent earned less than N100,000, (91) 28.6 percent earned between N100,000-N300,000, (54) 17.0 percent earned between N300,001-N500,000 and (17) 5.3 percent earned N500,001 and above. The diversity across respondents indicates that the data collected is suitable to achieve the aim of the study.

In response to how long they have been using e-channels, (36) 11.1 per cent responded 6 months to 1 year, (102) 32.1 per cent 1 to 5 years, (126) 39.6 per cent 6 to 10 years, while (54) 17 per cent above 10 years. This means that the majority of the respondents have been using e-channels for 1 year to 10 years. Concerning the e-channel that the respondents use for e-services, (318) 100 per cent use phone, Laptop (214) 67.3 per cent, Cybercafe (53) 16.7 per cent, POS (253) 79.6 per cent while ATM (280) 88.1 per cent. Most of the respondents use phones as the primary channel for e-services. This may be because the average Nigerian has a phone, and due to its convenience and mobility, it is used by all the respondents. In terms of the e-service they have been involved in, (202) 63.5 per cent have been involved in online shopping, (272) 85.5 per cent e-banking, (139) 43.7 per cent e-retailing, (176) 55.3 per cent e-ticketing, while (46) 14.5 per cent indicated others. This implies that the major e-service that the respondents are involved in is e-banking. This may be attributed to the current encouragement to use e-banking in Nigeria.

**Test of Reliability**

**TABLE 3**
**CRONBACH ALPHA COEFFICIENT (n=318)**

| Constructs | Number of Items | Cronbach Alpha Coefficient |
|---|---|---|
| Website Appearance | 6 | 0.803 |
| Ease of Use | 4 | 0.863 |
| Reliability | 5 | 0.799 |
| Security | 5 | 0.858 |
| Personalisation | 5 | 0.909 |
| Fulfilment | 4 | 0.766 |
| Responsiveness | 6 | 0.825 |
| Overall E-Service Quality | 6 | 0.842 |

Source: Survey (2021)

Table 3 shows the number of items that measured the variables in the study and the Cronbach Alpha Coefficients that were computed for the items that make up each construct. Cronbach's Alpha was used to assess the internal consistency reliability of the instrument. Hair, Black, Babin, Anderson and Tatham (2007) recommended an upper limit of 0.9. The Cronbach Alpha Coefficients in Table 3 range from 0.766 to 0.909. This indicates that the instrument is reliable and fit for use in the study.

**Relevant E-Service Quality Dimensions for E-Channels**

**TABLE 4**
**SUMMARY OF CORRELATION AND REGRESSION ANALYSIS OF E-SERVICE QUALITY DIMENSIONS AND OVERALL E-SERVICE QUALITY**

| Model | Correlation with Overall E-Service Quality | B | t-value | p-value |
|---|---|---|---|---|
| Constant | | -0.013 | -0.201 | 0.841 |
| Website Appearance | 0.484 | -0.003 | -0.136 | 0.892 |
| Ease of Use | 0.543 | 0.036 | 2.249 | **0.025** |
| Reliability | 0.915 | 0.674 | 32.643 | **0.000** |
| Security | 0.838 | 0.447 | 25.522 | **0.000** |
| Personalisation | 0.520 | 0.016 | 1.090 | 0.276 |
| Fulfilment | 0.560 | 0.045 | 2.443 | **0.015** |
| Responsiveness | 0.723 | -0.205 | -7.596 | **0.000** |
| **R** = 0.979 | | | | |
| **R$^2$** = 0.958 | | | | |
| **F-Value** = 1015.24 | | | | |
| **F-sig** = 0.000 | | | | |

Source: Survey (2021)

Table 4 shows the results of correlation analysis between the e-service quality dimensions and overall e-service quality, and the multiple regression results. The correlation between the e-service quality dimensions and overall e-service quality ranges from 0.484 to 0.915. Reliability (0.915), security (0.838) and responsiveness (0.723) dimensions are highly correlated with overall e-service quality. The Table also indicates that the model is statistically significant at 5% level of significance (F=1015.24, p=0.000<0.05). This indicates that there is a statistically significant relationship between e-service quality dimensions and overall e-service quality. This implies that website appearance, ease of use, reliability, security, personalisation, fulfilment, and responsiveness jointly determine overall e-service quality. The R Square value of 0.958 indicates the coefficient of determination. Therefore the e-service quality dimensions selected for this study explain 95.8 per cent of the variation in overall e-service quality. Ease of use, reliability, security and fulfilment dimensions were found to have a statistically significant positive influence on overall e-service quality, while responsiveness has a statistically significant negative influence. Website appearance and personalisation are not statistically significant. Comparatively, the dimensions of e-service quality that have significant influence on the perception of overall e-service quality include reliability ($b_3$=0.674, t=32.643, p=0.000), security ($b_4$=0.447, t=25.522, p=0.000), fulfilment ($b_6$=0.045, t=2.443, p=0.015), ease of use ($b_2$=0.036, t=2.249, p=0.025) and responsiveness ($b_7$= -0.205, t= -7.596, p=0.000) respectively.

Thus, the regression equation is:

$$Y = -0.013 + 0.036X_2 + 0.674X_3 + 0.447X_4 + 0.045X_6 - 0.205X_7 + e \qquad (2)$$

Where:
Y = Overall E-Service Quality; a = Constant; $b_2$, $b_3$, $b_4$, $b_6$, $b_7$= Regression Coefficients for $X_2$, $X_3$, $X_4$, $X_6$, $X_7$; $X_2$ = Ease of Use; $X_3$ = Reliability; $X_4$ = Security; $X_6$ = Fulfilment; $X_7$ = Responsiveness; e = error term

## DISCUSSION

The paper examined the measurement and dimensions of e-service quality and the relationship with overall e-service quality in the context of Nigeria. The study revealed that the introduction of e-services has led to the search and identification of dimensions suitable for e-channels. It has been noted that the dimensions of SERVQUAL cannot effectively measure e-service quality, so there is need to develop dimensions that are suitable for e-channels. Consequently, fifteen e-service quality dimensions were extracted from the literature. Based on the results of importance rating by e-channel users, seven e-service quality dimensions were selected and tested. Some researchers have identified these dimensions as determinants of overall e-service quality. The empirical validation of the seven e-service quality dimensions showed that only ease of use, reliability, security, fulfilment, and responsiveness dimensions statistically influence overall e-service quality. Website appearance and personalisation were not statistically significant.

The finding that website appearance is not significant is in disagreement with the results of some previous researchers that website appearance plays a vital role in how customers judge e-service quality (Cox & Dale, 2001; Santos, 2003; Lee & Lin, 2005; Hongxiu et al., 2009; Khushdil, 2018; Paulo et al., 2019). This implies that for e-channels, a simple website appearance that can allow e-channel users to perform the service is sufficient, as noted by Raman et al. (2008). The study found ease of use to be one of the major determinants of overall e-service quality, as in the study of Shirshendu & Sanjit (2011) and Narteh (2013; 2015). The results showed that reliability is the most significant determinant of overall e-service quality. This finding aligns with the results of Narteh (2013), Al-Hawary & Al-Smeran, (2017) and Tan et al. (2018).

In agreement with previous studies (Madu and Madu, 2002; Wolfinbarger and Gilly, 2003; Akinmayowa and Ogbeide, 2014; Paulo et al., 2019) security was found to be an important dimension of e-service quality. This implies that the security of e-channels influences the perception of overall e-service quality; however, Narteh (2013; 2015) did not found security to be important. This study did not find personalisation as a significant e-service quality dimension influencing overall e-service quality. This finding is at variance with the assertion of (Kaynama & Black, 2000; Poon & Lee, 2012) that personalisation influences customer behaviour.

The finding of this study that fulfilment has a significant influence on overall e-service quality agrees with the results of Sakhaei et al. (2014), Narteh (2015) and Paulo et al. (2019). In consonance with previous research (Madu and Madu, 2002; Narteh, 2015; Al-Hawary and Al-Smeran, 2017; Tan et al., 2018) that responsiveness is a critical factor in the perception of e-service quality, this study found responsiveness to be significant, but it has a negative influence on overall e-service quality. This implies that the attention and promptness of e-service providers in dealing with customer requests, questions, complaints and compensation of customers when they encounter challenges are essential to e-channel users. Still, it could have a negative influence on the perception of overall e-service quality. In as much as e-channel users expect that e-service providers should be responsive, when a customer experiences challenges with e-channels, it could negatively influence the perception of overall e-service quality.

## CONCLUSION

Service quality has been identified as one of the critical success factors in the management of services. However, the measurement appears to pose challenges to service firms and researchers because of the unique features of services. As such, several models and dimensions were propounded to measure and evaluate service quality. Initially, the models and dimensions were designed for direct services, where customers have face-to-face interaction with the business. Recently, the measurement of service quality of e-channels has become a significant issue facing business organizations. This study attempted to discuss the circumstances that led to the movement from traditional service quality to e-service quality and examined the avalanche of service quality dimensions that different researchers have used over the years. With this review, there is evidence that service quality is multi-dimensional and that the service quality dimensions of SERVQUAL are not sufficient to address e-service quality effectively. The paper went further to propose a model to measure the e-service quality dimensions for e-channels.

The information from this study has important managerial implications. Managers of online services should see e-service quality from the customer's perspective in order to meet or exceed their expectations. Based on the findings, there are five relevant e-service quality dimensions: reliability, security, fulfilment, ease of use and responsiveness for online channels. This is important for managers to identify the reasons for customers' perceived overall e-service quality and to address them satisfactorily. The results have provided insight to better understand the relevant e-service quality dimensions and the contribution of each dimension to overall e-service quality. These dimensions can provide practical levers for managers of online services to enhance customer service experience. Moreover, the dimensions identified in the study can provide managers with a direction to the essential factors to focus on to strengthen e-service delivery.